\documentclass{aa}
\usepackage{graphics}

\begin{document}
\thesaurus{02 (08.19.5 RX\,J1713.7-3946; 13.07.2)}
\title{Evidence for TeV gamma-ray emission from the shell type SNR
       RX\,J1713.7$-$3946}
\author{H.~Muraishi\inst{1},
T.~Tanimori\inst{2},
S.~Yanagita\inst{1},
T.~Yoshida\inst{1},
M.~Moriya\inst{2},
T.~Kifune\inst{3},
S.~A.~Dazeley\inst{4},
P.~G.~Edwards\inst{5},
S.~Gunji\inst{6},
S.~Hara\inst{2},
T.~Hara\inst{7},
A.~Kawachi\inst{3},
H.~Kubo\inst{2},
Y.~Matsubara\inst{8},
Y.~Mizumoto\inst{9},
M.~Mori\inst{3},
Y.~Muraki\inst{8},
T.~Naito\inst{7},
K.~Nishijima\inst{10},
J.~R.~Patterson\inst{4},
G.~P.~Rowell\inst{3,4},
T.~Sako\inst{8,11},
K.~Sakurazawa\inst{2},
R.~Susukita\inst{12},
T.~Tamura\inst{13},
and T.~Yoshikoshi\inst{3}
}

\authorrunning{H. Muraishi et al.}
\titlerunning{TeV gamma-ray emission from the SNR RX\,J1713.7$-$3946}
\offprints{yanagita@mito.ipc.ibaraki.ac.jp}

\institute{Faculty of Science, Ibaraki University,
   Mito, Ibaraki 310-8521, Japan
 \and Department of Physics, Tokyo Institute of Technology,
   Meguro, Tokyo 152-8551, Japan
 \and Institute for Cosmic Ray Research, University of Tokyo,
   Tanashi, Tokyo 188-8502, Japan
 \and Department of Physics and Mathematical Physics, University of
   Adelaide, South Australia 5005, Australia
 \and Institute of Space and Astronautical Science,
   Sagamihara, Kanagawa 229-8510, Japan
 \and Department of Physics, Yamagata University,
   Yamagata 990-8560, Japan
 \and Faculty of Management Information, Yamanashi Gakuin
   University,  Kofu, Yamanashi 400-8575, Japan
 \and STE Laboratory, Nagoya University,
   Nagoya, Aichi 464-8601, Japan
 \and National Astronomical Observatory, Tokyo 181-8588, Japan
 \and Department of Physics, Tokai University,
   Hiratsuka, Kanagawa 259-1292, Japan
 \and LPNHE, Ecole Polytechnique. Palaiseau CEDEX 91128, France
 \and Computational Science Laboratory, Institute of Physical
   and Chemical Research, Wako, Saitama 351-0198, Japan
 \and Faculty of Engineering, Kanagawa University,
   Yokohama, Kanagawa 221-8686, Japan}

\date{Received 1 December 1999 / Accepted 23 December 1999}

\maketitle

\begin{abstract}
We report the results of TeV gamma-ray observations of the shell type SNR
RX\,J1713.7$-$3946 (G347.3$-$0.5). The discovery of strong non-thermal X-ray
emission from the northwest part of the remnant strongly suggests the
existence of electrons with energies up to $\sim$ 100 TeV in the remnant,
making the SNR a good candidate TeV gamma-ray source. We observed
RX\,J1713.7$-$3946 from May to August 1998 with the CANGAROO 3.8m atmospheric
imaging \v Cerenkov telescope and obtained evidence for TeV gamma-ray emission
from the NW rim of the remnant with the significance of $\sim 5.6 \sigma$.
The observed TeV gamma-ray flux from the NW rim region was estimated to be
(5.3 $\pm$ 0.9[statistical] $\pm$ 1.6[systematic]) $\times$
10$^{-12}$ photons cm$^{-2}$ s$^{-1}$ at energies $\geq$ 1.8 $\pm$ 0.9 TeV.
The data indicate that the emitting region is much broader than the point
spread function of our telescope. The extent of the emission is consistent
with that of hard X-rays observed by ASCA. This TeV gamma-ray emission can be
attributed to the Inverse Compton scattering of the Cosmic Microwave
Background Radiation by shock accelerated ultra-relativistic electrons.
Under this assumption, a rather low magnetic field of $\sim11 \mu$G is deduced
for the remnant from our observation.

\keywords{gamma-rays: observations -- supernova remnant}

\end{abstract}

%________________________________________________________________

\section{Introduction}
Supernova remnants (SNRs) are currently believed to be a major source of
galactic cosmic rays (GCRs) from the arguments of energetics, shock
acceleration mechanisms (Blandford \& Eichler \cite{blandford87},
Jones \& Ellison \cite{jones91}), and the elemental abundances in the source
of GCR (Yanagita et al. \cite{yanagita90},
Yanagita \& Nomoto \cite{yanagita99}). EGRET observations suggest that the
acceleration sites of GCRs at GeV energies are SNRs (Esposito et al.
\cite{esposito96}). However, direct evidence for the SN origin of GCRs at TeV
energies is scarce (e.g.\ Koyama et al.\ \cite{koyama95},
Allen et al.\ \cite{allen97}, Buckley et al.\ \cite{buckley98}).
Arguably the best evidence for the existence of relativistic electrons with
energies around 100 TeV is the \,{CANGAROO} observation of TeV gamma-rays from
the northeast rim of SN1006, which coincides with the region of maximum flux
in the 2--10 keV band of the ASCA data (Tanimori et al. \cite{tanimori98b}).
This TeV gamma-ray emission was explained as arising from 2.7\,K Cosmic
Microwave Background Radiation (CMBR) photons being Inverse Compton (IC)
up-scattered by electrons with energies up to $\sim$ 100 TeV and allowed,
together with the observation of non-thermal radio and X-ray emission, the
estimation of the physical parameters of the remnant, such as the magnetic
field strength (Pohl \cite{pohl96}, Mastichiadis \cite{mastichiadis96},
Mastichiadis \& de Jager \cite{jager96}, Yoshida \& Yanagita \cite{yoshida97},
Naito et al. \cite{naito99}).

The shell type SNR RX\,J1713.7$-$3946 was discovered in the ROSAT All-Sky
Survey (Pfeffermann \& \,{Aschenbach} \cite{pfeffermann96}). The remnant has
a slightly elliptical shape with a maximum extent of $\sim 70\,^{\prime}$.
The 0.1--2.4 keV X-ray flux from the whole remnant is
$\sim$ 4.4 $\times$ 10$^{-10}$ erg cm$^{-2}$ s$^{-1}$
ranking it among the brightest galactic supernova remnants. Subsequent
observations of this remnant by the ASCA Galactic Plane Survey revealed strong
non-thermal hard X-ray emission from the northwest (NW) rim of the remnant
that is three times brighter than that from SN1006
(Koyama et al. \cite{koyama97}). The non-thermal emission from the NW rim
dominates the X-ray emission from RX\,J1713.7$-$3946, and the SNR X-ray
emission as a whole is dominated by non-thermal emission
(Slane et al. \cite{slane99}, Tomida \cite{tomida99}). It is notable that the
observed emission region of hard X-rays extends over an area
$\sim 0^{\circ}.4$ in diameter. Slane et al. (\cite{slane99}) carried out
843 MHz radio observations using the Molonglo Observatory Synthesis Telescope,
and discovered faint emission which extends along most of the SNR perimeter,
with the most distinct emission from the region bright in X-rays.
Slane et al. (\cite{slane99}) suggest the distance to RX\,J1713.7$-$3946 is
about 6 kpc based upon the observation of CO emission from molecular clouds
which are likely to be associated with the remnant.

The dominance of non-thermal emission from the shell is reminiscent of SN1006.
Koyama et al. (\cite{koyama97}) proposed from the global similarity of the new
remnant to SN1006 in its shell type morphology, the non-thermal nature of the
X-ray emission, and apparent lack of central engine like a pulsar, that
RX\,J1713.7$-$3946 is the second example, after SN1006, of synchrotron X-ray
radiation from a shell type SNR.
These findings from X-ray observations would suggest that TeV gamma-ray
emission could be expected, as observed in SN1006, from regions in the remnant
extended over an area larger than the point spread function of a typical
imaging telescope ($\sim 0^{\circ}$.2).

Both SN1006 and RX\,J1713.7$-$3946 show notably lower radio flux
densities and relatively lower matter densities in their ambient space when
compared with those for the other shell type SNRs (Green \cite{green98}) for
which the Whipple group (Buckley et al. \cite{buckley98}) and \,{CANGAROO}
group (Rowell et al. \cite{gavin99}) have reported upper limits to the TeV
gamma-ray emission. These characteristics might be related to the reason why
TeV gamma-rays have been detected only for SN1006 and not from other shell
type SNRs: the lower radio flux may indicate a weaker magnetic field which may
result in a higher electron energies due to reduced synchrotron losses.
In addition, the lower matter density would suppress the production of
$\pi^{0}$ decay gamma-rays. An observation of TeV gamma-rays from
RX\,J1713.7$-$3946 would provide not only further direct evidence for the
existence of very high energy electrons accelerated in the remnant but also
other important information on some physical parameters such as the strength
of the magnetic field which are relevant to the particle acceleration
phenomena occurring in the remnant, and would also help clarify the reason
why TeV gamma-rays have until now been detected only from SN1006.

With the above motivation, we have observed RX\,J1713.7$-$3946 with the
CANGAROO imaging TeV gamma-ray telescope in 1998. Here we report the result
of these observations.

\section{Instrument and Observation}
The CANGAROO 3.8m imaging TeV gamma-ray telescope is located near Woomera,
South Australia (136$^{\circ}$47'E, 31$^{\circ}$06'S)
(Hara et al. \cite{hara93}). A high resolution camera of
256 photomultiplier tubes (Hamamatsu R2248) is installed in the focal plane.
The field of view of each tube is about 0$^{\circ}$.12 $\times$
0$^{\circ}$.12, and the total field of view (FOV) of the camera is about
3$^{\circ}$. The pointing accuracy of the telescope is $\sim 0^{\circ}.02$,
determined from a study of the trajectories of stars of magnitude 5 to 6 in
the FOV. RX\,J1713.7$-$3946 was observed in May, June and August in 1998.
During on-source observations, the center of the FOV tracked the NW
rim (right ascension $17^{\rm h}\,11^{\rm m}\,56^{\rm s}.7$, declination
$-39^{\circ}\,31^{\prime}\,52^{\prime \prime}.4$ (J2000)), which is the
brightest point in the remnant in hard X-rays (Koyama et al. \cite{koyama97}).
An off-source region having the same declination as the on-source but a
different right ascension was observed before or after the on-source
observation for equal amounts of time each night under moonless and usually
clear sky conditions. The total observation time was 66 hours for on-source
data and 64 hours for off-source data. After rejecting data affected by
clouds, a total of 47.1305 hours for on-source data and
45.8778 hours for off-source data remained for this analysis.

\section{Analysis and Result}
The standard method of image analysis was applied for these data which is
based on the well-known parameterization of the elongated shape of the
\v Cerenkov light images using
``{\it width},''``{\it length},''``{\it concentration}'' (shape),
``{\it distance}'' (location), and the image orientation angle
\,{``{\it alpha}''} (Hillas \cite{hillas85}, Weekes et al. \cite{weekes89},
Reynolds et al. \cite{reynolds93}).
However, the emitting region of TeV gamma-rays in this target may be extended,
as in the case of SN1006. For extended sources, use of the same criteria as
for a point source in the shower image analysis is not necessarily optimal.
We made a careful Monte Carlo simulation for extended sources of various
extents and found the distribution of the shower image parameter of
{\it width}, {\it length}, and {\it concentration} for gamma-ray events is
essentially the same within a statistical fluctuation as in the case of
a point source. However, the simulation suggests that we should allow a wider
range dependent on the extent of the source for the parameter of
{\it distance} and {\it alpha} to avoid overcutting gamma-ray events.
In this analysis, gamma-ray--like events were selected with the criteria of
0.$^{\circ}$01 $\le$ {\it width} $\le$ 0.$^{\circ}$11,
0.$^{\circ}$1 $\le$ {\it length} $\le$ 0.$^{\circ}$45,
0.3 $\le$ {\it concentration} $\le$ 1.0 and
0.$^{\circ}$5 $\le$ {\it distance} $\le$ 1.$^{\circ}$2.

\begin{figure}[t]
\resizebox{8cm}{!}{\includegraphics{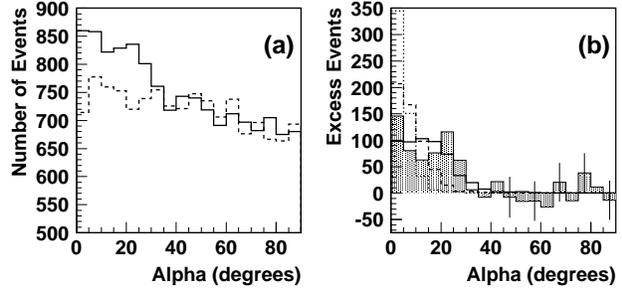}}
\caption
{(a) Distributions of the orientation angle ``{\it alpha}'' for gamma-ray-like
events with respect to the center of the field of view, which for on-source
data corresponds to the NW rim of RX\,J1713.7$-$3946. The solid line and
dashed line are for on-source and off-source data respectively.
(b) Distribution of the excess events of the on-source over the off-source
level shown in Figure~\ref{alpha}a, shown as the shaded bins. The vertical
bars for several bins indicate plus and minus one standard deviation which is
approximately the same for all bins. The expected {\it alpha} distribution for
a point source (dotted line), and disk-like sources with a radius of
0$^{\circ}$.2 (dashed line) and 0$^{\circ}$.4 (thick solid line) centered at
the FOV by the Monte Carlo method are also shown. Here the curves are
normalized to the actual excess number of gamma-ray-like events with
{\it alpha} $\le 30^{\circ}$.}
\label{alpha}
\end{figure}

Figure~\ref{alpha}a shows the resultant {\it alpha} distribution
when we analyzed the distribution centered at the tracking point
(right ascension $17^{\rm h}\,11^{\rm m}\,56^{\rm s}.7$, declination
$-39^{\circ}\,31^{\prime}\,52^{\prime \prime}.4$ (J2000)), which is the
brightest point in the remnant in hard X-rays (Koyama et al. \cite{koyama97}).
The solid line and the dashed line indicate the on-source and off-source data
respectively. Here we have normalized the off-source data to the on-source
data to take into account the difference in observation time and the variation
of trigger rates due to the difference in zenith angle between on- and
off-source data and due to subtle changes in weather conditions.
The value of the normalization factor $\beta$ is estimated to be 1.03 from
the difference in total obsevation time for on- and off-source measurements.
On the other hand, the actual value of
the normalization factor $\beta$ is estimated to be $\sim 0.99$ from the ratio
of $N_{\rm{on}}$/$N_{\rm off}$, where $N_{\rm on}$ and $N_{\rm off}$
indicate the total number of gamma-ray-like events with {\it alpha} between
40$^{\circ}$ and 90$^{\circ}$ for on- and off-source data respectively.
We selected the region with {\it alpha} $>$ 40$^{\circ}$ to avoid
any ``contamination'' by gamma-rays from the source, in the knowledge
that the source may be extended.
The small discrepancy in the two estimates of the value $\beta$ might
come from a slight change in the mirror reflectivity during the observation
due to dew. Here we adopt the value 0.99 for $\beta$ in the following
analysis by taking the small discrepancy into the systematic errors
due to the uncertainty in the mirror reflectivity as shown below.
Figure~\ref{alpha}b shows the {\it alpha}
distribution of the excess events for the on-source over the off-source
distribution shown in Figure~\ref{alpha}a. A rather broad but
significant peak can be seen at low {\it alpha}, extending to $ 30^{\circ}$.
The {\it alpha} distributions expected for a point source and several
disk-like extended sources of uniform surface brightness with various radii
centered at our FOV were calculated using the Monte Carlo method.
These distributions are shown in the same figure. The {\it alpha} distribution
of the observed excess events appears to favour a source radius of
$\sim 0^{\circ}.4$, which suggests the emitting region of TeV gamma-rays is
extended around the NW rim of RX\,J1713.7$-$3946.
The statistical significance of the excess is calculated by
($N_{\rm on}(\alpha) - \beta N_{\rm off}(\alpha)$) /
$\sqrt{N_{\rm on}(\alpha) + \beta^{2}N_{\rm off}(\alpha)}$,
where $N_{\rm on}(\alpha)$ and $N_{\rm off}(\alpha)$ are the numbers of
gamma-ray--like events with {\it alpha} less than $\alpha$ in the on- and
off-source data respectively.
The significance at the peak of the X-ray maximum was $5.6 \sigma$
when we chose a value of {\it alpha} $30^{\circ}$ considering
the result of the Monte Carlo simulation shown in Figure~\ref{alpha}b.

In order to verify this extended nature, we examined the effects of the cut
in shape parameters on the {\it alpha} distribution by varying each cut
parameter over wide ranges. We also produced {\it alpha} distributions for
different energy ranges and data sub-sets. Similar broad peaks in the
{\it alpha} distribution persisted through these examinations. Also we
examined more recent data from PSR1706$-$44 from July and August 1998
and obtained a narrow peak at {\it alpha} $< 15^{\circ}$, as expected for
a point source. This confirms that the extended nature of the TeV gamma-ray
emitting region does not come from some malfunction of our telescope system
and/or systematic errors in our data analysis. A similar, but not as broad,
{\it alpha} peak was seen for SN1006 (Tanimori et al. \cite{tanimori98b}).

\begin{figure}[t]
\resizebox{8cm}{!}{\includegraphics{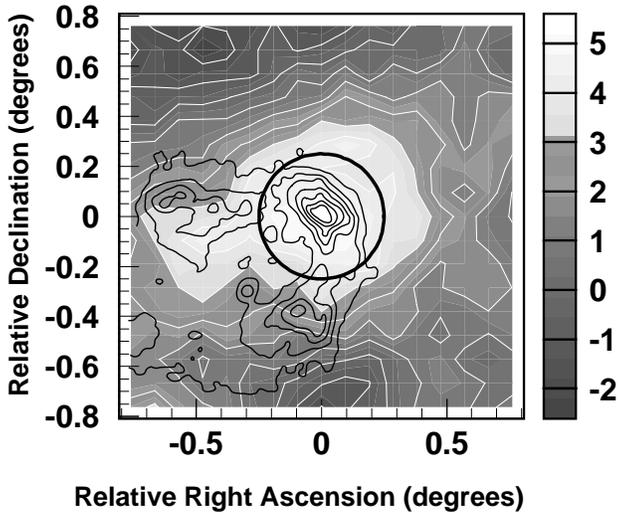}}
\caption
{Contour map of significance around the NW rim of RX\,J1713.7$-$3946 centered
at the region brightest in hard X-ray emission (right ascension
$17^{\rm h}\,11^{\rm m}\,56^{\rm s}.7$, declination
$-39^{\circ}\,31^{\prime}\,52^{\prime \prime}.4$ (J2000)). White lines
indicate the significance level. The contours of the 0.5--10 keV band of the
X-ray flux (from Tomida 1999) also are overlaid as solid lines.
The solid circle indicates the size of our PSF.}
\label{map-add}
\end{figure}

In order to see the extent of the emitting region, we made a significance map
of the excess events around the NW rim of RX\,J1713.7$-$3946.
Significances for {\it alpha} $\le 30^{\circ}$
were calculated at all grid points in $0^{\circ}.1$ steps in the FOV.
Figure~\ref{map-add} shows the resultant significance map
of the excess events around the NW rim of RX\,J1713.7$-$3946 plotted as
a function of right ascension and declination, in which the contours of the
hard X-ray flux (Tomida \cite{tomida99}) are overlaid as solid lines.
The solid circle indicates the size of the point spread function (PSF) of our
telescope which is estimated to have a standard deviation of
$\sim 0^{\circ}.25$ for {\it alpha} $\le 30^{\circ}$ based upon
Monte Carlo simulations for a point source with a Gaussian function.
The area which shows the highest significance in our TeV gamma-ray
observation coincides almost exactly with the brightest area in hard X-rays.
The region which shows the emission of TeV gamma-rays with high significance
($\ge 3 \sigma$ level) extends  wider than our PSF and appears to coincide
with the ridge of the NW rim that is bright in hard X-rays. It extends over
a region with a radius of $\sim0^{\circ}.4$. This region persisted in similar
maps calculated for several values of {\it alpha} narrower than $30^{\circ}$.

%-----------------------------------------------------------
%-----------------------------------------------------------

The integral flux of TeV gamma-rays was calculated, assuming emission from
a point source, to be
(5.3 $\pm$ 0.9 [statistical] $\pm$ 1.6 [systematic]) $\times$
10$^{-12}$ photons cm$^{-2}$ s$^{-1}$ ($\geq$ 1.8 $\pm$ 0.9 TeV).
The flux value and the statistical error were estimated from the excess number
of $N_{\rm on}(30^{\circ})-\beta N_{\rm off}(30^{\circ})$, where the value of
$30^{\circ}$ for {\it alpha} is chosen by the argument mentioned before.
The causes of the systematic errors are categorized by uncertainties in
(a) assumed differential spectral index,
(b) the loss of gamma-ray events due to the parameter cuts,
(c) the estimate of core distance of showers by the Monte Carlo method,
(d) the trigger condition,
(e) the conversion factor of the ADC counts to the number of photo-electrons,
and (f) the reflectivity of the reflector.
These errors from (a) to (f) are estimated as
15\%, 22\%, 3\%, 12\%, 10\%, and 8\% for the integral flux and
24\%, 2\%, 8\%, 20\%, 29\%, and 17\% for the threshold energy, respectively.
The total systematic errors shown above are obtained by adding those errors
quadratically.

To summarise, all our observed data support the hypothesis that the
emitting region of the NW rim is extended.
In general, the value of the effective detection area of the telescope system
for extended sources would be reduced by some factor from that for a point
source, because the gamma-ray detection efficiency decreases with the distance
of emitting points from the center of the FOV when we observe with a single
dish. We calculated the efficiency as a function of the distance by the Monte
Carlo method by analyzing the data with the same criteria as applied to the
actual data. We estimated the value of the correction factor to the effective
area to be $\sim 1.2$ for our target by integrating the efficiency over the
distance for an extended disk-like source of uniform surface brightness with
a radius of $0^{\circ}.4$. The factor of 1.2 is less significant than the
systematic errors estimated above.

%-----------------------------------------------------------
%-----------------------------------------------------------

\section{Discussion}
The SNR RX\,J1713.7$-$3946 is reminiscent of SN1006 both in the synchrotron
X-ray emission from the shell far from the centre of the remnant and also in
the TeV gamma-ray emission from an extended region coincident with that of
the non-thermal X-rays. This suggests that the particles responsible for the
emission of the high energy photons are accelerated in shocks.

There are several possible emission processes of TeV gamma-rays:
the emission induced by accelerated protons (by the $\pi^{0}$ decay process)
and by electrons -- through bremsstrahlung and/or the Inverse Compton (IC)
process. The expected integral flux of gamma-rays above our threshold energy
of $\sim 1.8$ TeV by the $\pi^{0}$ decay process is estimated to be
$< 4 \times 10^{-14}$ photons cm$^{-2}$ s$^{-1}$ (Drury et al. \cite{drury94},
Naito \& Takahara \cite{naito94}), where we assume the distance and the upper
limit for the number density in the ambient space of the remnant as 6 kpc and
0.28 atoms/cm$^{3}$, respectively (Slane et al. \cite{slane99}).
This flux value is too low to explain our observed flux, even taking into
account the large uncertainties in the estimates of the distance and
the ambient matter density of the remnant (Slane et al. \cite{slane99},
Tomida \cite{tomida99}). However, there remains the possibility of some
contribution of the $\pi^{0}$ decay process if the remnant is interacting
with a molecular cloud located near the NW rim (Slane et al. \cite{slane99}).
The relative contribution in emissivity of the bremsstrahlung
process compared to $\pi^{0}$ decay process is estimated as $\sim 10$ \%,
assuming the flux ratio of electrons to protons is $\sim 1/100$ and that both
have power law spectra with the index of 2.4 (Gaisser \cite{gaisser90}),
indicating this process is also unlikely to dominate. Therefore, the most
likely process for TeV gamma-ray emission seems to be the IC process.

Under this assumption, the magnetic field strength in the supernova remnant
can be deduced from the relation
$L_{\rm syn}/L_{\rm IC} = U_{\rm B}/U_{\rm ph}$ between
the IC luminosity $L_{\rm IC}$ and synchrotron luminosity $L_{\rm syn}$,
where $U_{\rm B} = B^{2}/8\pi$ and $U_{\rm ph}$ are the energy densities
of the magnetic field and the target photon field, respectively.
$L_{\rm syn}$ and $L_{\rm IC}$ in the above formula must be due to electrons
in the same energy range. The value of $L_{\rm syn}$ which should be compared
with our TeV gamma-ray data is estimated from the ASCA result to be
$L_{\rm syn}=L_{\rm ASCA}\int_{E^{\rm min}_{\rm syn}}^{\infty}E^{-1.44}dE$
/$\int_{\rm 0.5keV}^{\rm 10keV}E^{-1.44}dE$, extrapolating
the synchrotron spectrum with the same power law out of the energy range of
0.5$-$10 keV covered by ASCA (Tomida \cite{tomida99}). Here
$L_{\rm ASCA}=2.0\times10^{-10}$ erg cm$^{-2}$ s$^{-1}$ is the X-ray
luminosity in the 0.5$-$10 keV energy band observed by ASCA from the NW rim
of the remnant and the power law index of $-1.44$ is the mean value for
index of X-rays in the same energy range (Tomida \cite{tomida99}).
$E^{\rm min}_{\rm syn}=0.14(B/10 \mu \mbox{G})$ keV is a typical synchrotron
photon energy emitted by electrons which emit 1.8 TeV photons (the threshold
energy of our observation) by the IC process when we assume the target photons
to be from the CMBR. The value of $L_{\rm IC}$ is
calculated to be $4.2 \times 10^{-11}$ erg cm$^{-2}$ s$^{-1}$ from our result
for the number of photons of TeV gamma-rays, and using the fact that the
spectra of synchrotron photons and IC photons follow the same power law when
the electrons have a power law spectrum. Thus inserting
$L_{\rm syn}$, $L_{\rm IC}$, and $U_{\rm ph}=4.2\times10^{-13}$ erg cm$^{-3}$
of the energy density for the CMBR into the above relation, we can solve for
the magnetic field strength $B$. Finally, the magnetic field at the NW rim
is estimated to be $\sim 10.8 \mu$G. The extrapolation used to estimate
$L_{\rm syn}$ is reasonable, because $E^{\rm min}_{\rm syn}$ is
estimated to be 0.15 keV; this is not so different from the minimum energy of
the ASCA band (0.5 keV).

The electrons responsible for the synchrotron and IC photon emissions
are likely to have been accelerated by the shocks in the remnant as
discussed above. If the maximum electron energy is limited by
synchrotron losses, this maximum energy can be estimated by equating
the cooling time due to synchrotron losses with the time scale of
acceleration by the first Fermi process in a strong shock as $\sim 50
(V_{\rm s}/2000 \mbox{km s$^{-1}$}) (B/10 \mu \mbox{G})^{-0.5}$ TeV,
where V$_{\rm s}$ is the shock velocity (Yoshida \& Yanagita
\cite{yoshida97}). On the other hand, equating the acceleration
time with the age of the remnant, the maximum energy can be expressed
$\sim 180 (V_{\rm s}/2000 \mbox{km s$^{-1}$})^{2} (B/10 \mu \mbox{G})
(t_{\rm age}/10000 \mbox{year})$ TeV.  In either case, whether it is
synchrotron losses or the age of the remnant that limits the maximum
electron energy (Reynolds \& Keohane \cite{reynolds99}),
electrons should exist with energies high enough to emit the observed
synchrotron X-rays and TeV gamma-rays by the IC process.

It is notable that both RX\,J1713.7$-$3946 and SN1006 have relatively low
magnetic field strengths and low matter densities in their ambient space.
These common features may have arisen if the magnetic field was `frozen in'
to the matter without amplification other than by compression by shocks and
may be the reason why electrons are accelerated to such high energies.
These facts may also explain the radio quietness (Green \cite{green98}) and
the weak emissivity of $\pi^{0}$ decay gamma-rays of the remnants.
For SN1006, the low matter density in the ambient space might result from
the remnant being located far off the galactic plane and the supernova being
of type Ia. For RX\,J1713.7$-$3946, the low matter density may be caused by
material having been swept out by the stellar wind of the supernova progenitor
(Slane et al. \cite{slane99}).
The low magnetic field and the low matter density in the ambient space of
SN1006 and RX\,J1713.7$-$3946 may explain why TeV gamma-rays have been
detected so far only for these two remnants.

In conclusion, we have found evidence for TeV gamma-ray emission from
RX\,J1713.7$-$3946 at the level of 5.6 sigma.
If confirmed (\`a la Weekes \cite{weekes99}), this would be the second case
after SN1006 to show directly that particles are  accelerated up to energies
of $\sim 100$ TeV in the shell type SNR.

\begin{acknowledgements}

We sincerely thank H.\ Tomida and K.\ Koyama for providing us the ASCA data.
We thank the referee very much for his helpful comments on the paper.
This work is supported by a Grant-in-Aid for Scientific Research from Japan's
Ministry of Education, Science, and Culture, a grant from the Australian
Research Council and the (Australian) National Committee for Astronomy (Major
National Research Facilities Program), and the Sumitomo Foundation.
The receipt of JSPS Research Fellowships (SH, AK, GPR, KS, and TY) is also
acknowledged.

\end{acknowledgements}

\end{document}